\documentclass[conference]{IEEEtran}
\usepackage{cite}
\usepackage{graphicx}
\usepackage{float}
\graphicspath{ {figs/} }
\usepackage{amsmath}
\usepackage[ruled,vlined]{algorithm2e}
\usepackage{xcolor}
\usepackage[font=small]{caption}
\usepackage{subcaption}
\usepackage{siunitx}
\usepackage{colortbl}
\usepackage{algorithmic}
\usepackage[thinlines]{easytable}
\usepackage{geometry}
\usepackage[normalem]{ulem}
\def\BibTeX{{\rm B\kern-.05em{\sc i\kern-.025em b}\kern-.08em
    T\kern-.1667em\lower.7ex\hbox{E}\kern-.125emX}}

\begin{document}
\title{Machine Learning on Camera Images for Fast mmWave Beamforming}

\author{
    \author{
    \IEEEauthorblockN{
        Batool Salehi,
        Mauro Belgiovine,
        Sara Garcia Sanchez,
        Jennifer Dy,
        Stratis Ioannidis, and
        Kaushik Chowdhury
    }
    \IEEEauthorblockA{
        Department of Electrical and Computer Engineering\\
        Northeastern University, Boston, USA\\
        Email: \{salehihikouei.b,belgiovine.m,garciasanchez.s\}@northeastern.edu, \{jdy,ioannidis,krc\}@ece.neu.edu
    }
}
}

\maketitle
\begin{abstract}
Perfect alignment in chosen beam sectors at both transmit- and receive-nodes is required for beamforming in mmWave bands. Current 802.11ad WiFi and emerging 5G cellular standards spend up to several milliseconds exploring different sector combinations to identify the beam pair with the highest SNR. In this paper, we propose a machine learning (ML) approach with two sequential convolutional neural networks (CNN) that uses out-of-band information, in the form of camera images, to (i) rapidly identify the locations of the transmitter and receiver nodes, and then (ii) return the optimal beam pair. We experimentally validate this intriguing concept for indoor settings using the NI 60GHz mmwave transceiver. Our results reveal that our ML approach reduces beamforming related exploration time by 93\% under different ambient lighting conditions, with an error of less than 1\% compared to the time-intensive deterministic method defined by the current standards. 
\end{abstract}

\begin{IEEEkeywords}
Millimeter wave (mmWave), out-of-band beamforming, machine learning, image processing, location inference.
\end{IEEEkeywords}
\section{Introduction}
The looming spectrum crunch caused by billions of connected devices as well as the escalating demand for wireless resources to support high data rate, real-time multimedia content has resulted in immense interest in using mmWave frequencies for communication. Emerging 5G standards are poised to leverage frequencies in the 24--100GHz range within the mmWave band, thus assuring  multi-gigabit downlink data rate for users \cite{6732923}. However, since communication links in this band attenuate rapidly, transmitters generally use phased arrays with beamforming, so as to concentrate the electromagnetic energy in a narrow aperture~\cite{roh2014millimeter}. Hence, mmWave links must be formed with optimal alignment of the beams between the transceiver pair to be effective. Indeed, this first step consumes 
up to several seconds in current WiFi standards. Thus, for widespread deployment in time-critical applications, we propose a radically different approach that uses camera images as input to a two-stage CNN for guiding the beam selection process.

\subsection{Need for Beamforming in mmWave Links}
  \begin{figure}
    \centering
    \includegraphics[width=0.44\textwidth]{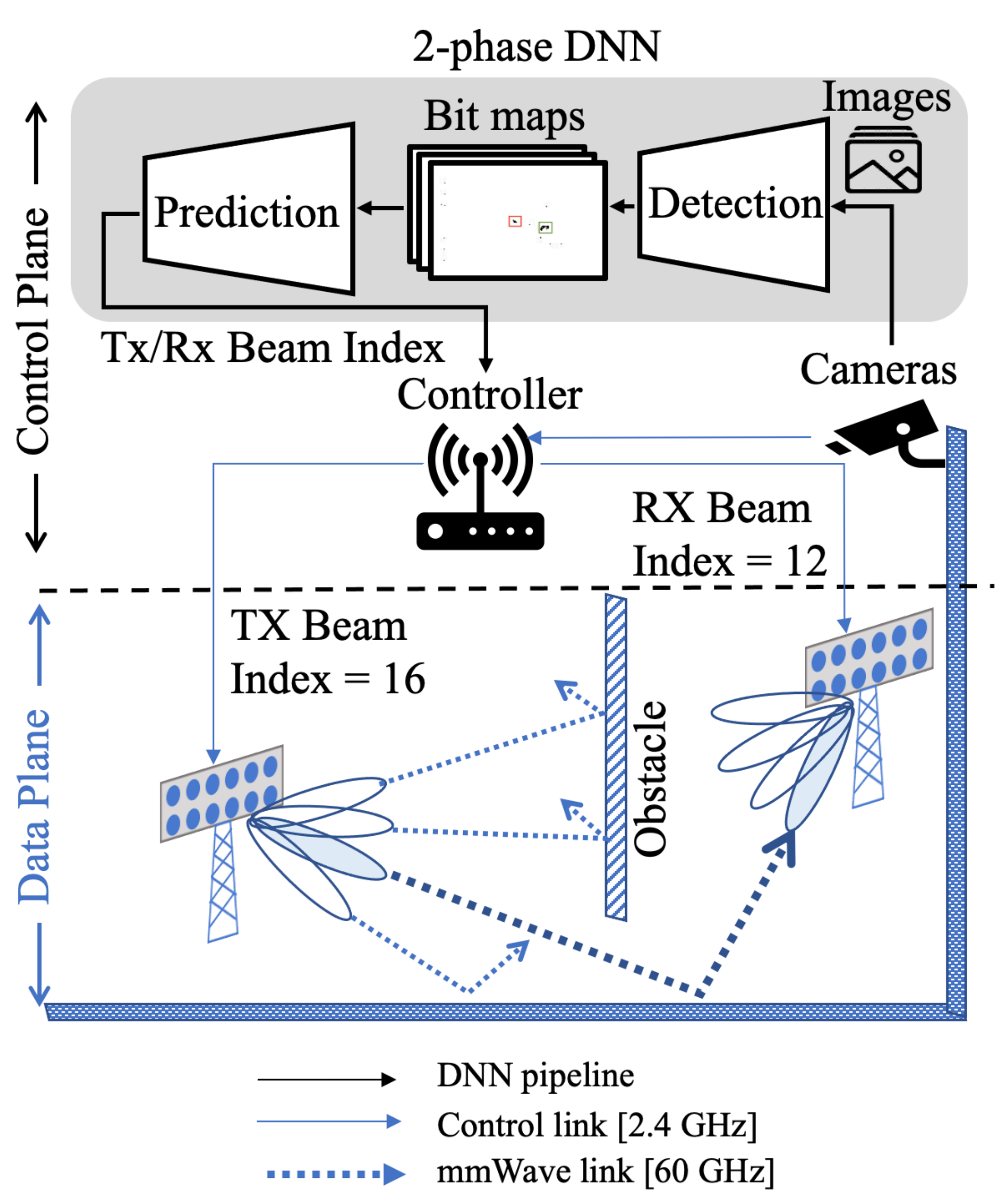}
    \caption{A camera observes users to find the best beam pair configuration for data transmission. The images pass through two stages, Detection and Prediction, in our pipeline, and the inferred best beam pairs are sent to the users by the network controller.}
    \label{use case}
\end{figure}

While narrow beams are better suited to combat the  atmospheric absorption and low penetration aspects of mmWave links, highly directional transmissions require an exhaustive search among different candidate beam orientations, concisely represented as a \textit{codebook}. Advanced phased arrays promise codebooks 
in 3-dimensions with up to 64 sectors per phased array, according to the 802.11ad standard \cite{assasa2019enhancing}, which further complicates a sequential search among all beam options. 

Current mmWave standards incorporate the following method for beam selection via the so called \textit{beam sweeping procedure}:  Different pairs of transmitter-receiver beams within a known codebook are successively chosen, and their performance in terms of signal strength is evaluated to determine the best pair for communication. For COTS 802.11ad routers, this process takes at least tens of ms \cite{yaman2016reducing}, an order of magnitude above the 1ms maximum latency required by the 5G Ultra-Reliable Low Latency Communications (URLLC) \cite{hou2019prediction}. Moreover, in a dynamic scenario, beam sweeping must be  periodically repeated in order to ensure directional links. Every time beam sweeping is performed, this action disrupts ongoing communication.  
\subsection{Scenario Description}

Given the undesirable delay arising from the standard beamforming procedure, we propose to leverage visual information as a potential solution to mitigate the beam training overhead. A schematic of our proposed approach is demonstrated in Fig.~\ref{use case}. The control unit gets visual snapshots of the environment taken by single/multiple cameras as input and directly predicts the best beam configuration at both transmitter and receiver ends that maximizes the SNR at the receiver. 

The acquired visual information passes through our two-stage pipeline in the Control Plane, as depicted in Fig.~\ref{use case}. In the first stage, \textit{Detection}, a deep convolutional neural network generates bit maps to indicate the relative transmitter and receiver location. The bit maps are then used in the second stage, \textit{Prediction}, to infer the best beam configuration. Finally, the best beam pair predictions, (16,12) in the figure, are sent to the transmitter and receiver by the network controller. 

Our approach does not require any hardware modifications at the user end. We specifically focus on indoor, dynamic, and rich multipath environments, such as offices, that typically suffer from high Non-Line-of-Sights (NLoS) probability. Notice that in this type of scenarios, the presence of obstacles in the Line-of-Sight (LoS) path causes certain beam pairs to achieve the highest performance among all evaluated pairs, through reflections on walls or certain surfaces.
\subsection{Proposed Approach}
We use machine learning to estimate the best beam pair based on input images. Our proposed method consists of a set of two sequential CNNs and can be summarized as follows:
\begin{itemize}
    \item \textbf{Stage 1:} We locate the transmitter and receiver radios in the input images and discard non-relevant information. In order to do that, we design a binary classifier trained to classify each portion of the incoming image, taken from our testbed on various light conditions, as either \textit{Antenna array} or \textit{Background}. We then create a quantized version of the input image by dividing it into small crops and classifying them individually, arranging the binary decision output of each crop in order to obtain a 2D bit map. 
    
    \item \textbf{Stage 2:} 
    We use a second CNN that accepts bitmaps obtained from the previous stage as input and predict the best beam configuration index at both transmitter and receiver. After predicting the best configuration pair, the corresponding beam weights are extracted from the codebook table.
    
\end{itemize} 

\subsection{Summary of Contributions}
Our main contributions are as follows:
\begin{itemize}
    \item We investigate the technical requirements for using visual information to boost beam selection operation in the 60 GHz mmWave band. Then, we propose a two-stage deep CNN architecture to properly map input images to the best beam configuration which maximizes the SNR at the receiver. Our proposed method achieves up to $99\%$ accuracy on best beam pair prediction in low light conditions.
    \item We design a testbed to validate our proposed method using National Instruments mmWave Transceiver \cite{niwebsite}. 
    To the best of our knowledge, this is the first work that experimentally validates the beam selection approach using visual information. All of the current beam prediction literature are based on synthetic data driven by ray tracing software. Such softwares for professional use come with expensive licenses, and those that are freely available may not consider side lobes and scattering/reflection from the surrounding environments, which limits use in real-world scenarios. 
    \item We configure our setup to support simultaneous beam alignment between transmitter and receiver. Moreover, we demonstrate that our proposed approach outperforms the exhaustive beam sweeping with 93\% reduction in the time required for beam initialization.
\end{itemize}
\section{Related Work}

      

\begin{figure}
    \centering
    \includegraphics[width=0.9\linewidth]{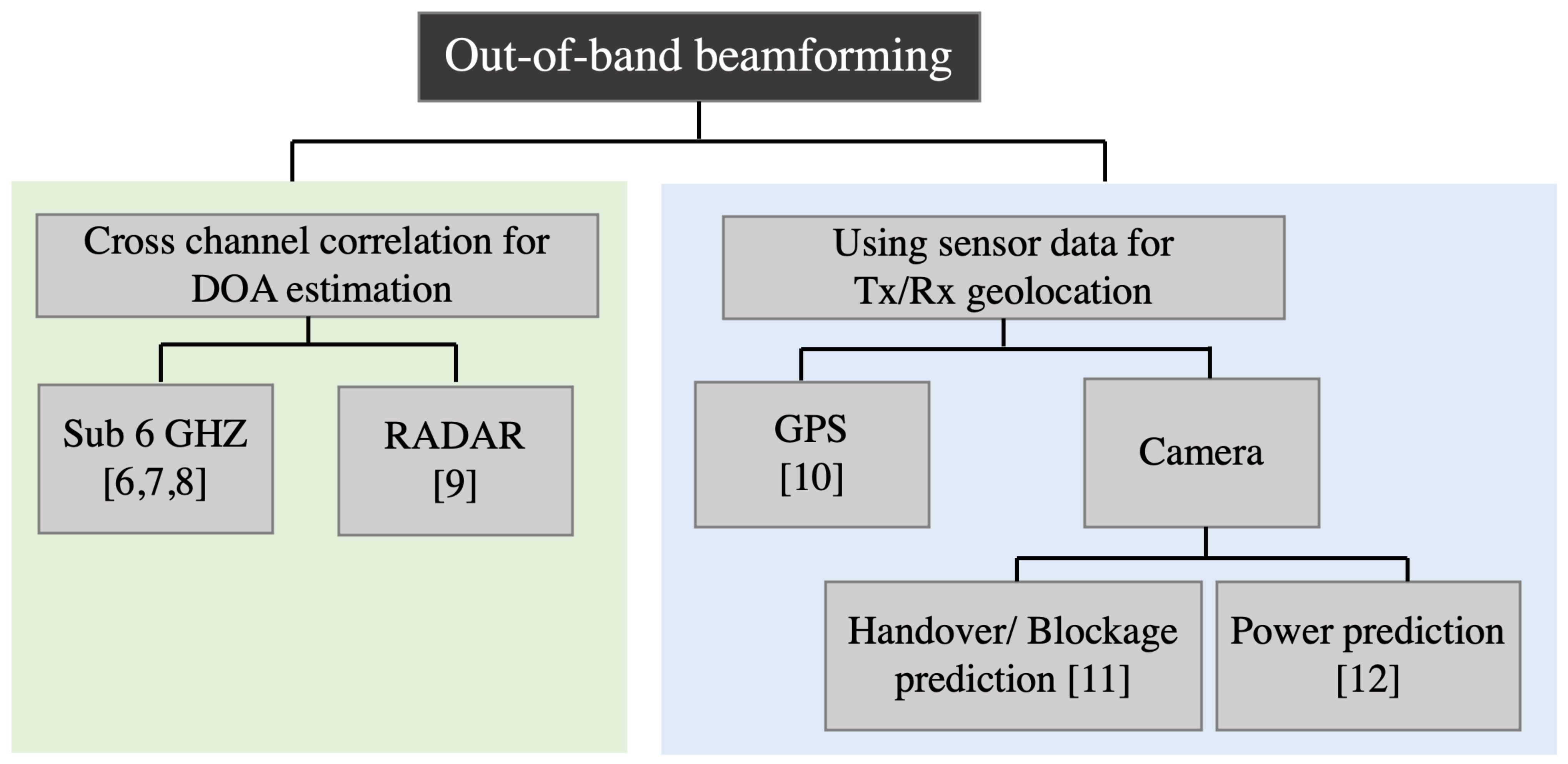}
    \caption{Classification in visual information aided beamforming.}
    \label{fig:related_work}
\end{figure}

In this section, we review the out-of-band methods for guiding beam sweeping, as they are the most comparable to our proposed method. Fig.~\ref{fig:related_work} shows a classification diagram of out-of-band beamforming methods.

\subsection{Cross Channel Correlation for DOA Estimation:}
This method attempts to reduce the beam sectors for searching by establishing a mapping between the channel measurements in the mmWave band and other frequencies.
\begin{itemize}
    \item \textbf{Sub-6GHz band:} In \cite{8101513} the spatial correlations with sub-6 GHz and mmWave band signals is used to speed up the initial beam alignment process. Using MUSIC algorithm, the AoA is estimated in the sub-6GHz, and the exhaustive search runs only for angles in range $A_{sub-6} \pm 10$ in the mmWave band. Steering with eyes closed \cite{nitsche2015steering} exploits the omni-directional transmissions at low frequencies to infer the LoS direction between the communicating devices to speed up the mmWave sector selection. Anum et al. \cite{ali2017millimeter} incorporate sub-6GHz bands in the form of a weighted sparse recovery approach with structured random codebooks to reduce the beam sweeping delay. 
    \item \textbf{RADAR band:} 
    \cite{gonzalez2016radar} shows that the main DoAs for the radar signal at 76.5 GHz and the mmWave signals at 65 GHz are comparable. As a result, the RADAR signals can be used to estimate the covariance of the received signal and channel information.
\end{itemize}

\subsection{Sensor Data for Tx/Rx Geolocation:} 
Knowing the geographical location of the transmitter and receiver can speed up the detection of best beam sector.
\begin{itemize}
    \item \textbf{GPS:} There are several works on using GPS to speed up the beam selection process\cite{va2017inverse}. MmWave beam prediction with situational awareness \cite{wang2018mmwave} uses the geographical information derived from a synthetic environment with moving vehicles to predicts the received beam power for each codebook element. We note that GPS does not work in indoor environments. Furthermore, the extracted locations need to be very precise and also include the orientation of the antenna, which is not provided by conventional GPS.
    \item \textbf{Camera:} The existing literature on image driven beamforming can be categorized into two parts: 
    \begin{enumerate}
        \item Hand-over among multi base stations by blockage prediction: In \cite{alkhateeb2018machine} a scenario with a single user and multiple base stations is considered. The base stations use the previous observations to predict blockage on a certain link in the next few frames. This allows the serving base stations to proactively hand-over the user to another base station in case of pending blockage.
        \item Estimate power in the Next Time Slot: \cite{8792137} proposes an approach to predict the time series of the received power at the receiver end. The transmitter and receiver are fixed, and a human, modeled as a cylinder, blocks the line-of-sight path. The sequential images are generated and labeled with received power in several hundred milliseconds ahead and fed to a neural network to predict the received power.
    \end{enumerate}
\end{itemize}

\subsection{Other Works Leveraging ML in Beamforming}
 In \cite{8395149} a mobile user is served by a number of distributed yet coordinating BSs. The user sends $N_{tr}$ pilots using an omni-directional antenna. Every BS switches between its legible beam patterns in the codebook and calculates the achievable rate of each direction. A deep neural network is then trained to maximize the cumulative data rate. In other words, the received signal is used as a signature to estimate the location of the user. 
\section{Experimental Setup and Dataset collection}
\label{sec:sec3}
We construct a testbed to examine the performance of our proposed method on a real dataset. In this section, we explain our approach for designing the experiment, collecting the dataset, and creating the  data processing pipeline. First, we discuss the beam sweeping latency measured from two different mmWave hardware in Section \ref{sec:Beam-sweepingLatency}. We describe National Instruments mmWave Transceiver, with 2GHz bandwidth at 60 GHz frequency, in Section\ref{sec:NIinstruments}. Then, in Section \ref{sec:expSetup}, we thoroughly describe implemention of our testbed, including the experiment setup description. In Section \ref{sec:datasetCollection}, we explain our approach for collecting data and the parameter used to evaluate the Quality of Link (QoL). Finally, we present an illustration of our dataset structure in Section \ref{preprocessinf and description}.
\subsection{Beam Sweeping Latency }
\label{sec:Beam-sweepingLatency}
In Table \ref{compare different devices}, we provide 
the measured beam sweeping time 
from 
two mmWave hardware. In particular, we consider the Terragraph channel sounders \cite{Terragraph}, a customized pair of nodes from Facebook designed for the channel modeling of 60GHz links, and the National Instruments mmWave Transceiver, that we use in our experiments. 
From table \ref{compare different devices}, we notice that the delay for establishing a link is in the order of milliseconds, due to the beam sweeping procedure. Moreover, values presented in Table \ref{compare different devices} only consider a fraction of the actual complete beam sweeping time. This fraction corresponds to the less time-consuming refinement stage, which assumes limited knowledge on the relative position between transmitter and receiver, bounded within a certain angular sector. However, in the WiFi standard 802.11ad, the complete beam sweeping procedure can take up to tens of seconds. 

\begin{table}
\caption{Exhaustive beam sweeping time for two mmWave hardware}
\centering
\begin{tabular}{ |c|c|c| } 
 \hline
 Device & Terragraph & NI \\ 
 \hline
 \hline
 Possible codebook configurations & 40  & 625 \\ 
 \hline
 Beam sweeping delay  & 11 ms  & 12.5 s\\ 
 \hline
 Beam sweeping delay per beam pair & 0.2750 ms  & 20 ms\\ 
 \hline
\end{tabular}
\label{compare different devices}
\end{table}

\subsection{National Instruments mmWave Transceiver}
\label{sec:NIinstruments}
For data collection, we use the mmWave transceiver system from National Instruments that supports real-time over the air mmWave communication. It operates in the 60 GHz frequency band with a bandwidth of 2GHZ. It is comprised of PXIe (PCI extensions for Instrumentation) chassis, controllers, a clock distribution module, FPGA modules, high-speed ultra wide-band DACs and ADCs, and LO and IF modules. The NI mmWave transceiver is implemented using seven FPGAs, each of which is responsible for an operation, such as coding, modulation, etc. Modules are controlled and synchronized using a central FPGA equipped with LabView software. It supports a variety of modulation schemes from BPSK to 16QAM, alongside with turbo coding. After being processed by FPGA, the signal is converted using DAC and sent over the air using RF front ends. 

\begin{figure}
    \centering
    \begin{subfigure}{0.29\textwidth}
        \centering
        \includegraphics[scale=0.31]{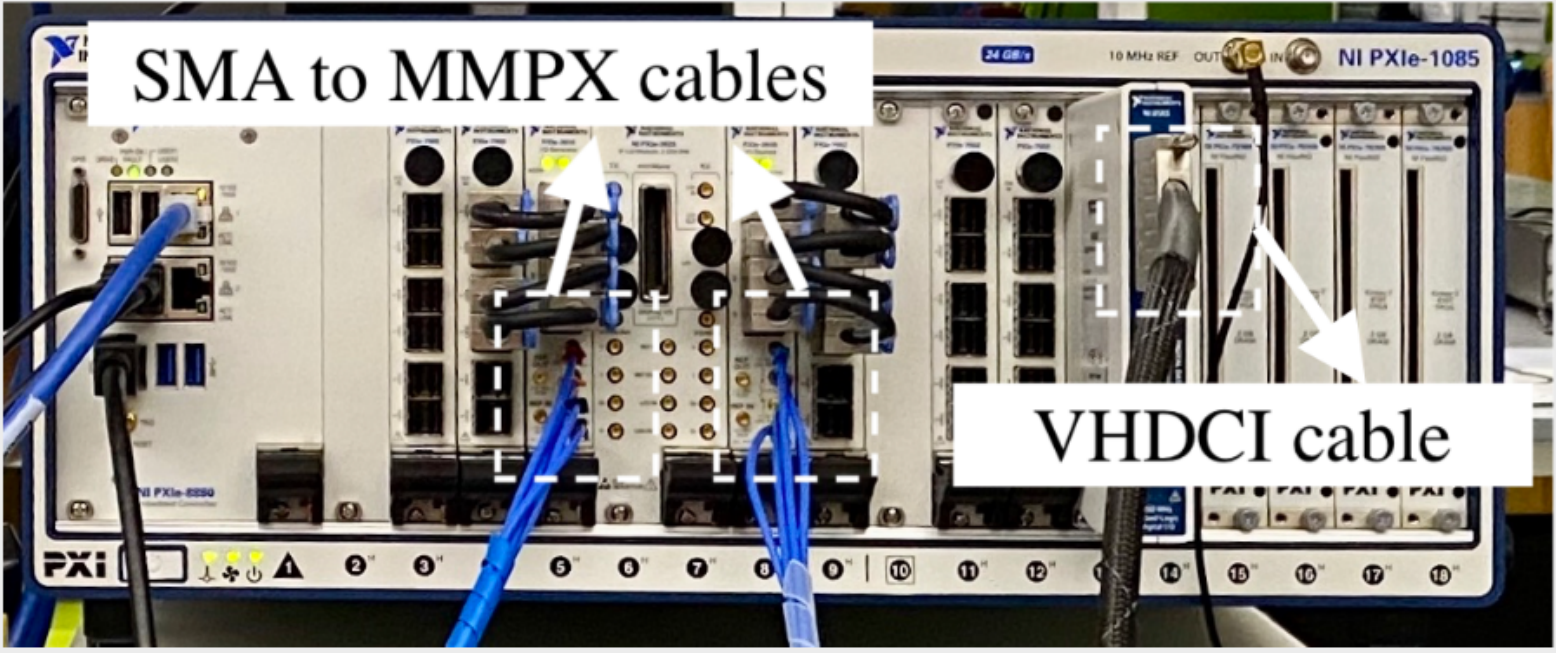}
        \caption{NI radio front panel}
        \label{Ni panel}
    \end{subfigure}
    \begin{subfigure}{0.17\textwidth}
        \centering
        \includegraphics[scale=0.123]{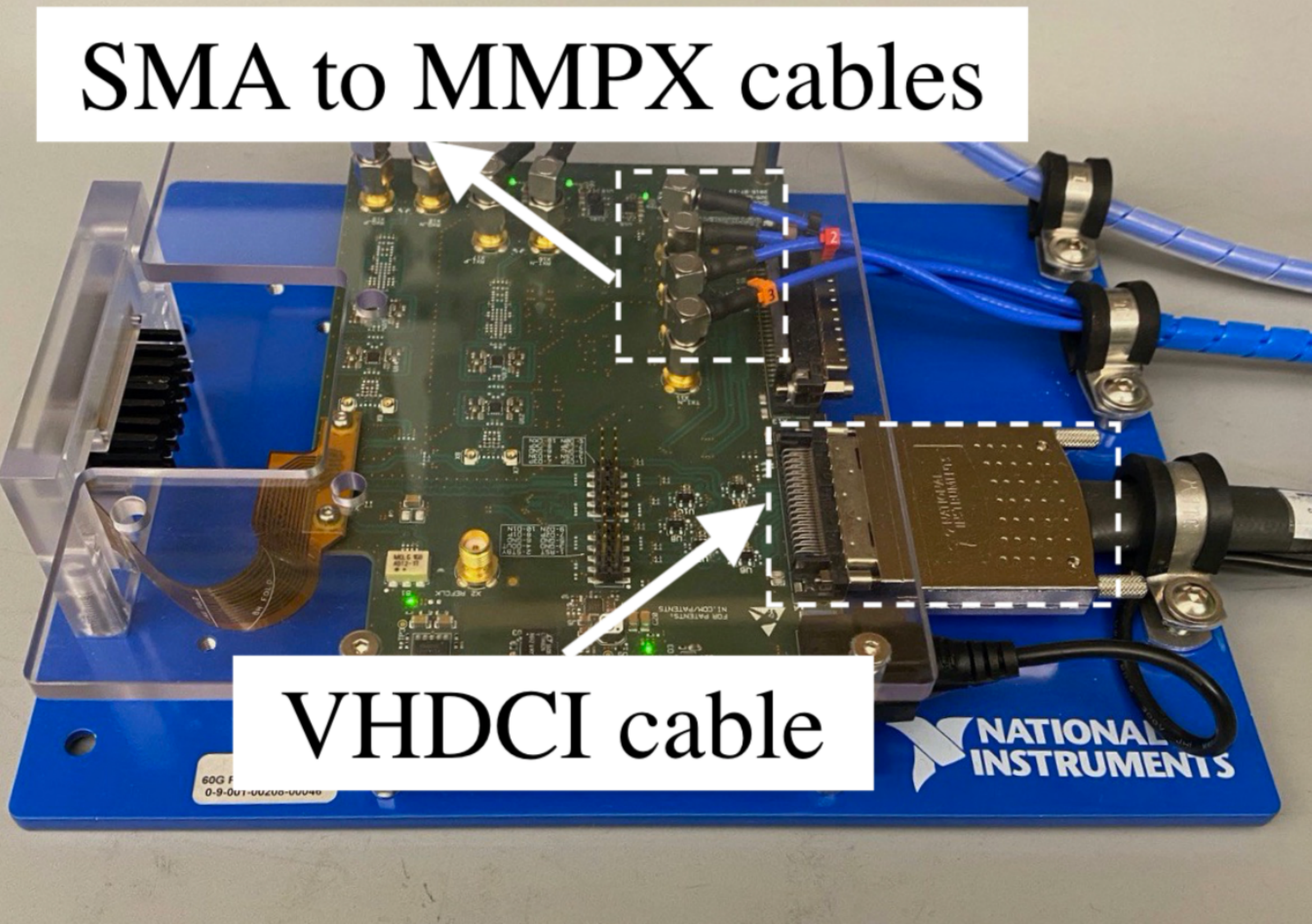}
        \caption{Sibeam connections}
        \label{Sibeam-connections}
    \end{subfigure}
    \caption{The NI radio front panel and Sibeam phased antenna array head. The SMA to MMPX cables are used for transferring I/Q samples, while VHDCI cables pass the control commands, including transmission direction.}
    \label{NI radio and Sibeam}
\end{figure}

In our experiment, we use SiBeam RF heads, a phased antenna array with 24 radiating elements. Each radiating element consists of a squared patch antenna of dimension $\SI{0.1}{\centi\metre}$. Among the 24 antenna elements, half of them are used for transmission, and the remaining for reception. The transmit power of each element is 1 dBm, resulting in a total transmit power of 12 dBm. The Sibeam antenna array supports only-azimuth beam sweeping as well as 2D  beam-sweeping in azimuth and elevation. The azimuth codebook includes 25 beams designed to horizontally sweep angles from ${-60}^{\circ}$ to ${+60}^{\circ}$ with an angular resolution-separation between two consecutive beams of ${5}^{\circ}$ and 3dB-beamwidth of ${25}^{\circ}$. Moreover, the 2D codebook has a total of 11 beams, sweeping azimuth and elevation angles. Fig.~\ref{NI radio and Sibeam} shows the front panel of NI radio and Sibeam antenna head. The radio and antenna head are connected by VHDCI and SMA to MMPX cables that are used for implementing LabView code and passing I/Q samples to antenna heads, respectively.

\subsection{Experimental Setup}
\label{sec:expSetup}
The testbed is deployed in a room of size of $\SI{310}{\centi\metre}\times\SI{510}{\centi\metre}$. The transmitter and receiver are mounted on a mobile slider each. These mobile sliders laterally move in the horizontal direction with a range of $\SI{120}{\centi\metre}$. The sliders are lifted up from the ground by $\SI{1}{\metre}$, and the distance between two sliders is fixed at $\SI{350}{\centi\metre}$. The movement speed and the stop time of the sliders are programmed using a controller. A red box is bonded on the top of the antenna array, making it distinctive in the image. 

An obstacle is located in between the sliders, blocking the LoS path between transmitter and receiver in certain directions. We collect the dataset for two types of obstacles, wood and card box. The obstacles are rectangular with dimensions $\SI{33}{\centi\metre}\times\SI{88}{\centi\metre}\times\SI{3}{\centi\metre}$ and $\SI{33}{\centi\metre}\times\SI{88}{\centi\metre}\times\SI{10}{\centi\metre}$ for wood and card box, causing 30dB and 4dB attenuation while blocking the LOS path, respectively. 

\begin{figure}
    \centering
    \includegraphics[scale=0.32]{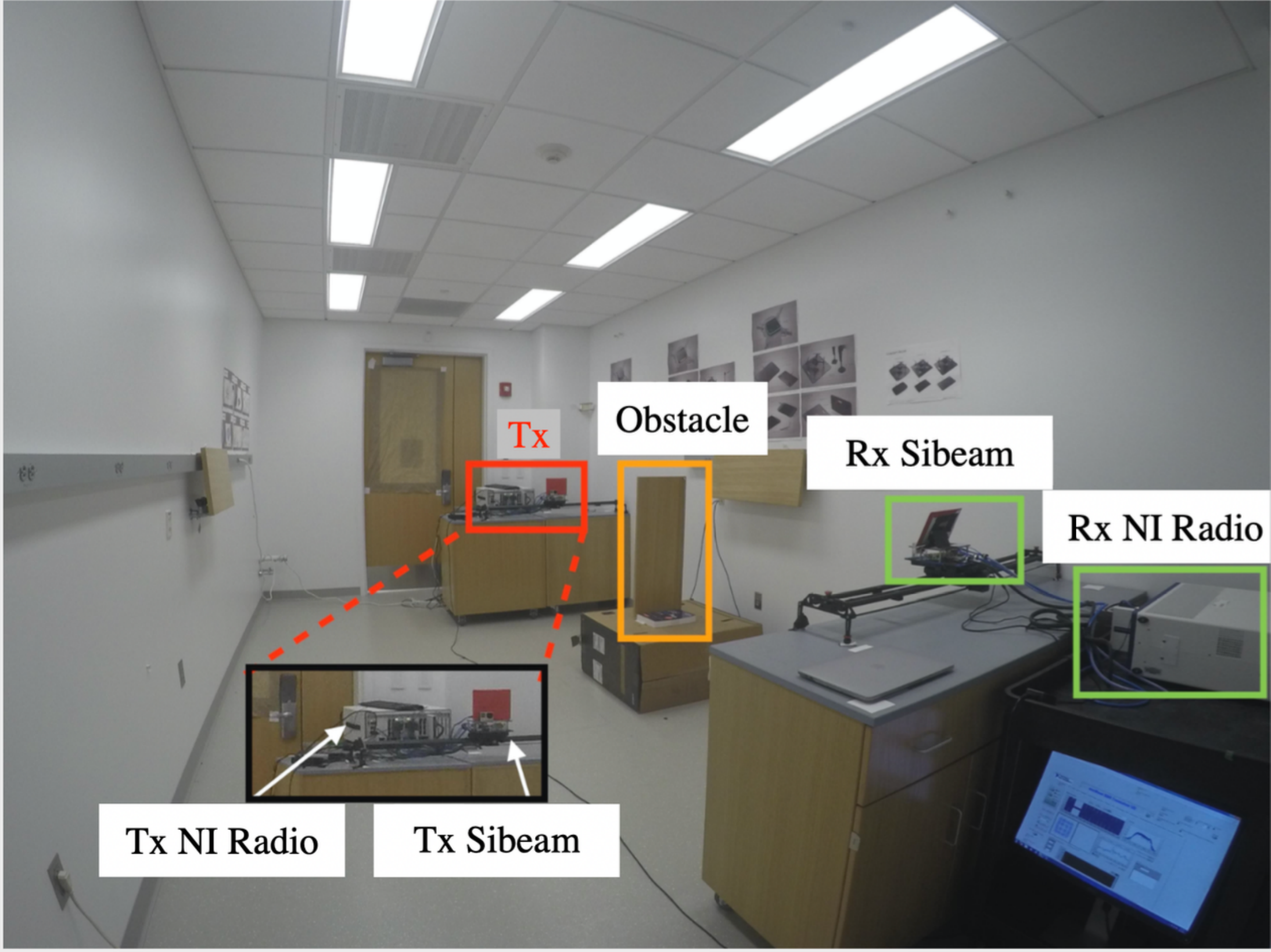}
    \caption{Testbed setup from the first camera angle perspective.}
    \label{testbed}
\end{figure}

\begin{figure}
    \centering
    \includegraphics[scale=0.38]{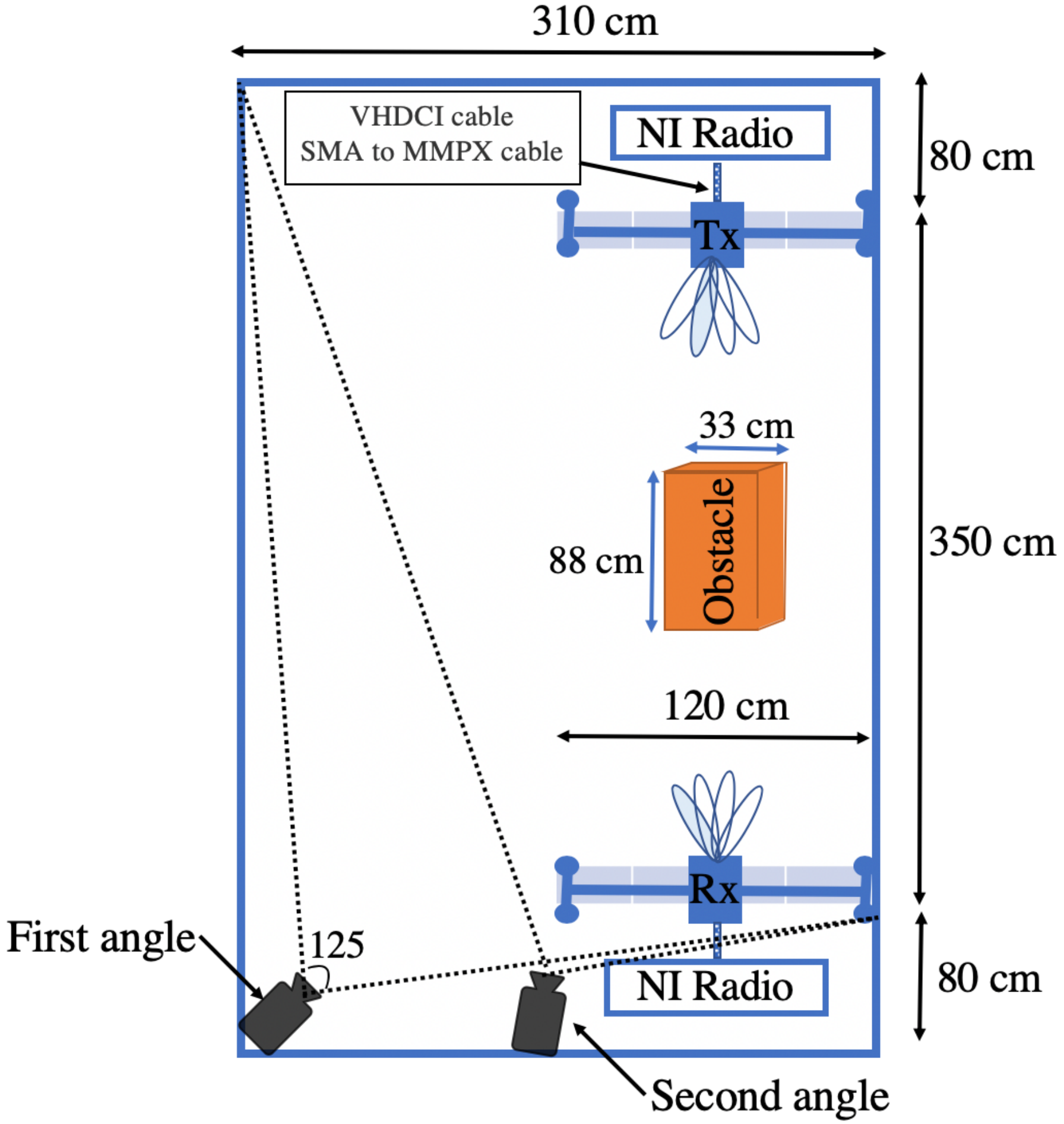}
    \caption{A diagram representing experiment parameters. The transmitter and receiver are mounted on two sliders and move in the horizontal direction while an obstacle blocks the LOS path in some cases. The cameras are located at the same height with different view angles. }
    \label{diagram testbed}
\end{figure}

Two GoPro Hero 4 cameras placed on at the height $\SI{169}{\centi\metre}$ from the ground monitor the movements in the room. The resolution of the cameras is 12Mp with FOV of ${125}^{\circ}$. The first angle, as shown in Fig.~\ref{diagram testbed} has a clear perspective on the transmitter and receiver. 
On the contrary, the second angle cannot see the transmitter in some cases, as the obstacle blocks the view. Fig.~\ref{testbed} shows the experiment setup from the camera perspective for the wooden obstacle from the first angle. In Fig.~\ref{testbed}, the antenna array inside of the red and green boxes are the transmitter and receiver, respectively. Fig.~\ref{diagram testbed} depicts a diagram from the testbed with experiment setting parameters.

\subsection{Dataset Collection}
\label{sec:datasetCollection}
For simplicity and ease of data collection, we consider 5 discrete, equally separated positions along the slider length. As a result, the gap between two consecutive locations is $\SI{24}{\centi\metre}$. This results in 25 distinct configurations for the transmitter and receiver locations. Each configuration is identified by a pair representing the relative location of the transmitter and receiver. We refer to the set of possible locations as the case set, $\{(i,j)| i=1,...,5 ~,~ j=1,...,5\}$.
For instance, case $(3,3)$ (as shown in Fig.~\ref{testbed}) is associated with a scenario in which the transmitter and receiver are both located at the third point from the wall. 

We choose the azimuth codebook as our reference since the beam switching is more tangible in one direction. Furthermore, we reduce the codebook size to 13 beams, by dropping the odd beam indexes from the default codebook. As a result, in order to perform beam sweeping at both transmitter and receiver sides, a total of 169 pairs of beams need to be evaluated for each case to determine the best one. 

\begin{figure}
    \centering
    \includegraphics[scale=0.47]{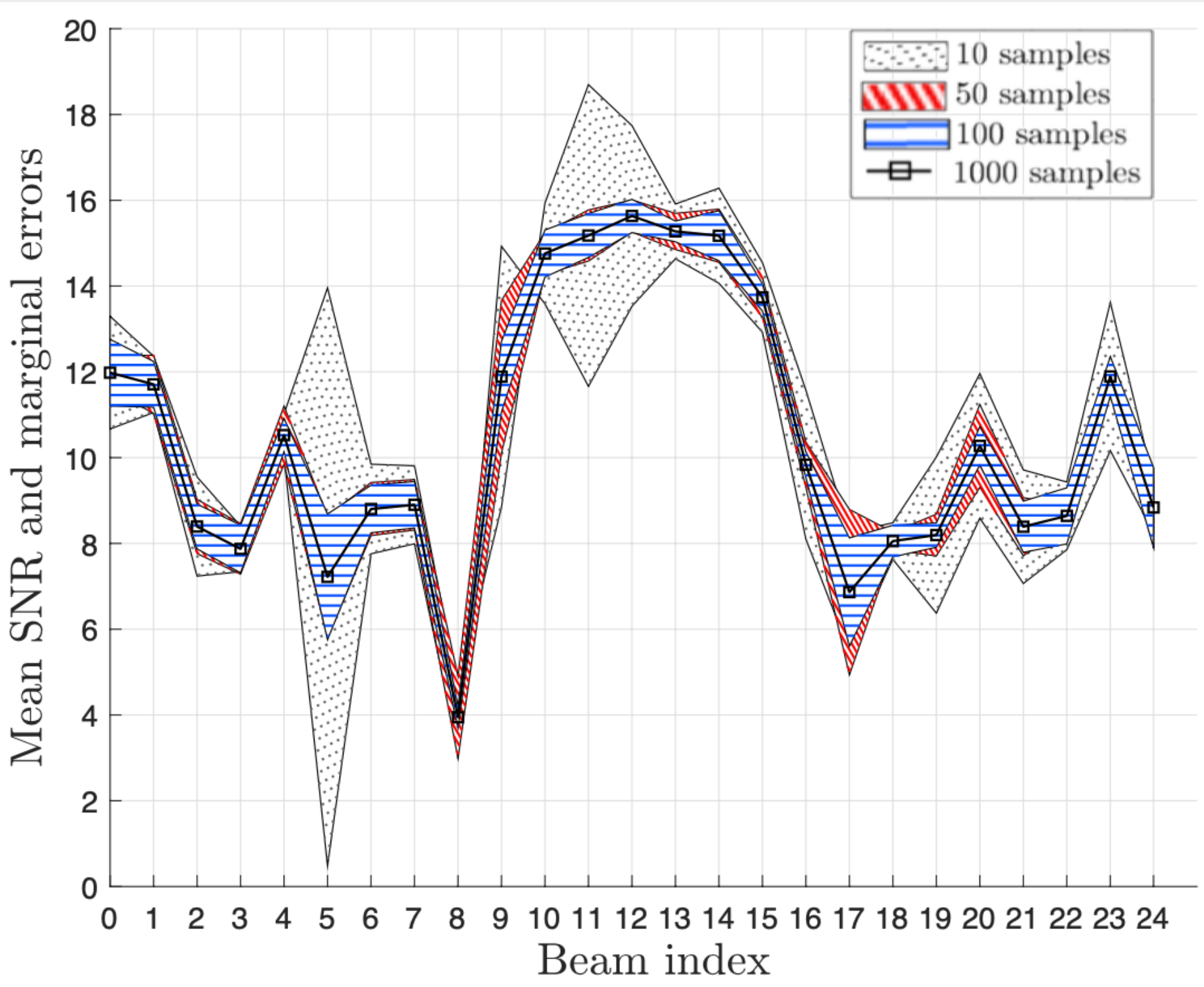}
    \caption{Marginal SNR error for three number of samples. The black line shows our reference, the mean SNR over 1000 samples.}
    \label{sample}
\end{figure}

We use the received SNR as our metric to evaluate the link quality. For each case, we collect a certain number of samples $N$ for all possible beam configurations. To determine $N$, we run a simple experiment: we fix the transmitter beam index to be 12, which corresponds to the antenna broadside direction (perpendicular direction to the axis containing the slider). Then, we sweep all possible beam indexes at the receiver and record the SNR for 1000 samples that we use as the reference. In Fig.~\ref{sample}, the black line shows the mean SNR of the reference, while each color depicts the marginal error in the logarithmic scale for three different sample numbers.
We select $N=50$ as the number of samples to be captured per beam pair, since increasing the number of samples does not contribute to an immense increase in measurement accuracy. The mean absolute error for 50 samples is $0.1077~dB$ over all codebook elements at the receiver.

We repeat the experiment for both obstacles, wood and card box. For the wooden obstacle, we capture one image per case, first angle in Fig.~\ref{testbed}. For the card box, we take two images per case from both first and second angles. We use this dataset to explore the effect of blocked viewpoints in section \ref{Results:blocked view}.

\subsection{Preprocessing and Dataset Description}
\label{preprocessinf and description}
The NI mmWave transceiver reports SNR as NaN (Not-a-Number) when the received power is lower than a threshold (-48 dB). In order to incorporate this in our QoL assessment, while processing the measured SNR samples, we interpret NaN values as a case of severe attenuation causing connection loss. 
\\We denote the codebook of possible beam configurations at the transmitter and receiver by $C_{Tx}$ and $C_{Rx}$ defined as:
\begin{equation}
    C_{Tx}=\{t_1,\dots,t_M\},~~  C_{Rx}=\{r_1,\dots,r_N\},
\end{equation}
where $M,N$ are the number of transmitter and receiver codebook elements, respectively. We define the set of possible beam configurations as:
\begin{equation}
    S =\{(t_{m},r_{n})|t_{m}\in C_{Tx},r_{n}\in C_{Rx}\}.
\end{equation}
With $|S|=M\times N $, recall that the transmitter and receiver need to sweep through all beam pair configurations in order to discover the best one. For a specific beam configuration $(t_{m},r_{n})\in S$, we define our quality metric $Q_{t_{m},r_{n}}$ as follows:
\begin{equation}
     Q_{t_{m},r_{n}} = \frac{E\left[SNR_{k}\right]}{N_{null}+1},\quad k = 1,...,K,
     \label{quality}
\end{equation}
where $K$ is the total number of valid SNR values, $E$ represents the mean operator and $N_{null}$ is the number of NaN values appeared while collecting data for beam pair $(t_{m},r_{n})$. Using \eqref{quality}, we assess the link quality of  every discrete device positioning $(i,j)$, with transmitter at location $i$ and receiver at location $j$, in order to select the best beam index pair $(t^*,r^*)\in S$. The result of this process is a set:
\begin{equation}
    B = \{b_{(t^*,r^*)}^{ij}\}\quad (t^*,r^*)\in S
\end{equation}

where each element is an ordered pair defined as: 
\begin{equation}
\label{eq:tuple}
b_{(t^*,r^*)}^{ij} = ~\langle(i,j),(t^*,r^*)\rangle
\end{equation}
The first elements in \eqref{eq:tuple} denote devices' positions $(i,j)$ and the second element is the associated best beam configuration, obtained as follows:
\begin{equation}
    \label{eq:best_beam_pair}
     (t^*,r^*) =  \underset{1\leq m \leq M , 1\leq n \leq N}{\arg\max}~(Q_{t_{m},r_{n}}).
\end{equation}

Table \ref{Tab:dataset} represents the dataset structure for both obstacles. For instance, from this table, we observe that the best beam pair for case $(i,j) = (3,1)$, i.e. the case in which the transmitter is at the third point and receiver is at the first point, is $(t^*,r^*) = (10,24)$ for wood and $(t^*,r^*) = (8,10)$ for card box as the obstacle. 
The dataset contains 25 different cases. Recall that we want our beam configuration estimator to be robust to light variations, while the other elements of the environment remain static. Therefore, we augment our dataset by applying 50 different light conditions, ranging from darker to lighter versions of the original sample, on each image in the dataset, resulting in 1250 training samples total. 

\begin{table}[t!]
\begin{center}
\begin{tabular}{|c|c|c|}
\hline
Case&\multicolumn{2}{c|}{Beam angles $(t^*,r^*)$} \\
\cline{2-3} 
(i,j) & Wood& Card box\\
\hline
 (1,1) & (14,14) & (12,12)\\  
 (1,2) & (14,14) & (14,14)\\
 (1,3) & (16,14) & (14,14)\\
 (1,4) & (18,16) & (16,14)\\
 (1,5) & (18,24) & (18,18)\\
 (2,1) & (10,10) & (12,14)\\
 (2,2) & (8,12) & (14,12)\\
 (2,3) & (14,14) & (14,14)\\
 (2,4) & (8,16) & (16,14)\\
 (2,5) & (16,16) & (16,12)\\
\rowcolor{lightgray} (3,1) & (10,24) & (8,10)\\
 (3,2) & (12,8) & (14,14)\\
 (3,3) & (10,12) & (14,14)\\
 (3,4) & (12,14) & (16,12)\\
 (3,5) & (16,12) & (6,14)\\
 (4,1) & (8,8) &  (8,8)\\
 (4,2) & (10,8) & (8,10)\\
 (4,3) & (10,8) & (14,12)\\
 (4,4) & (12,24) & (6,12)\\
 (4,5) & (12,0) & (22,12)\\
 (5,1) & (8,8) & (8,6)\\
 (5,2) & (8,8) & (12,10)\\
 (5,3) & (8,10) & (8,8)\\
 (5,4) & (12,8) & (12,8)\\
 (5,5) & (12,0) & (14,14)\\
 \hline
\end{tabular}
\caption{Best beam pair for two types of obstacles, wood and card box. We use \eqref{eq:best_beam_pair} to derive best beam pair based on our measurements.}
\label{Tab:dataset}
\end{center}
\end{table}
\section{Proposed Method}
\begin{figure}
    \centering
    \includegraphics[scale=0.33]{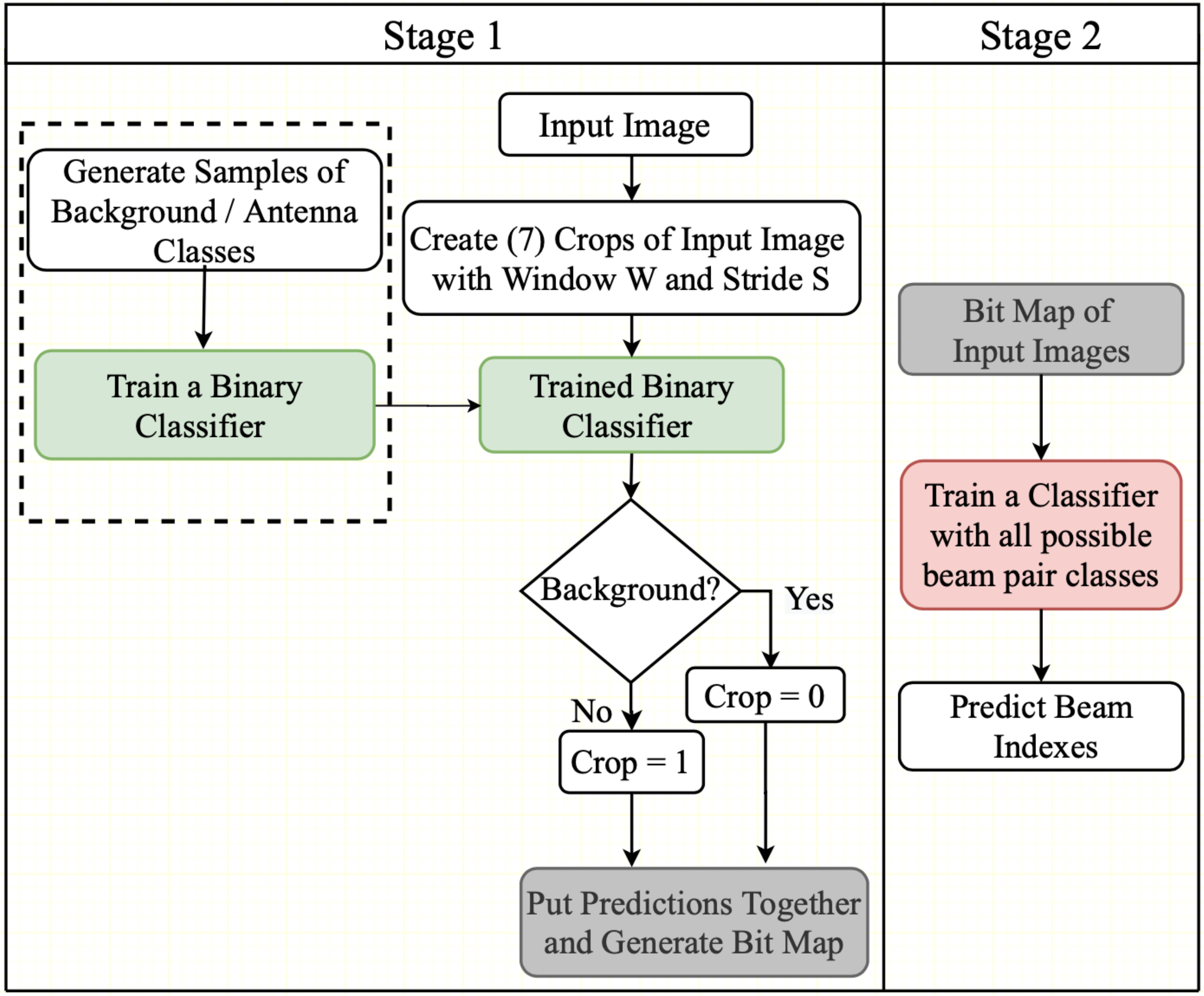}
    \caption{
    An overview of our two-stage CNN for fast beam prediction. In Stage 1, we infer the locations of transmitter and receiver devices in the input image, by dividing the image in crops and assigning them either a \textit{Background} (0) or \textit{Antenna array} (1) label. We then arrange the predictions into a 2D \textit{bit map} that we input to the second CNN model. In Stage 2, the generated bit maps from Stage 1 along with best beam pairs are used to train a second CNN.
    }
    \label{diagram}
\end{figure}

In this section, we present our two-stage CNN for finding the best beam index pair based on input images.
Fig.~\ref{diagram} summarizes our proposed pipeline for fast beam alignment using images.

\subsection{Stage 1: Inferring Transmitter and Receiver Locations }
\label{extract relevant information}
In our experiment, all the elements in the room are static except for the transmitter and receiver. Consequently, we conclude that the main cause of best beam variations is the relative movement between them. In Stage 1, we carefully infer the location of transmitter and receiver devices in the input image. In contrary to a simple image classification approach (i.e. by treating every pixel information in the picture as relevant to our task), our approach tries to identify the portions of the image that represent more relevant features, in this case, the antenna arrays positions.

We design and train a binary classifier with two outputs, namely \textit{Background}, which corresponds to non-relevant portions, and \textit{Antenna array}. To construct the training dataset for the binary classifier, from each input image, we create a set of windowed image crops having size $W\times W$ pixels. Starting from the upper left pixel, after generating the first crop the window moves by a step of S pixels, referred to as stride size, until the entire image is swept. Each crop is labeled as \textit{Background} or \textit{Antenna array}. If the input image has the shape of $H\times L$, then each image is reduced to a certain number of crops, according to the following equation:
\begin{equation}
    N_{crops} =\lfloor\frac{H-W}{S}+1\rfloor \times \lfloor\frac{L-W}{S}+1\rfloor.
    \label{crops}
\end{equation}

Since the antenna arrays comprise only a small portion of the image, we expect to have more samples for the \textit{Background} rather than the \textit{Antenna array} class. In order to obtain model robust to the light variations and obtain a balanced dataset, we exploit data augmentation by (1) applying different light conditions on fly while generating the training dataset and (2) keeping multiple copies of \textit{Antenna array} class input samples under different light conditions until we reach the same number of samples as \textit{Background} class. We split our dataset as $(70\%,15\%,15\%)$ for train, validation and test sets, respectively, and train a CNN binary classifier on the generated $W\times W$ input samples and relative labels, i.e. \textit{Antenna array} and \textit{Background}. 

\begin{figure}
    \centering
    \includegraphics[scale=0.35]{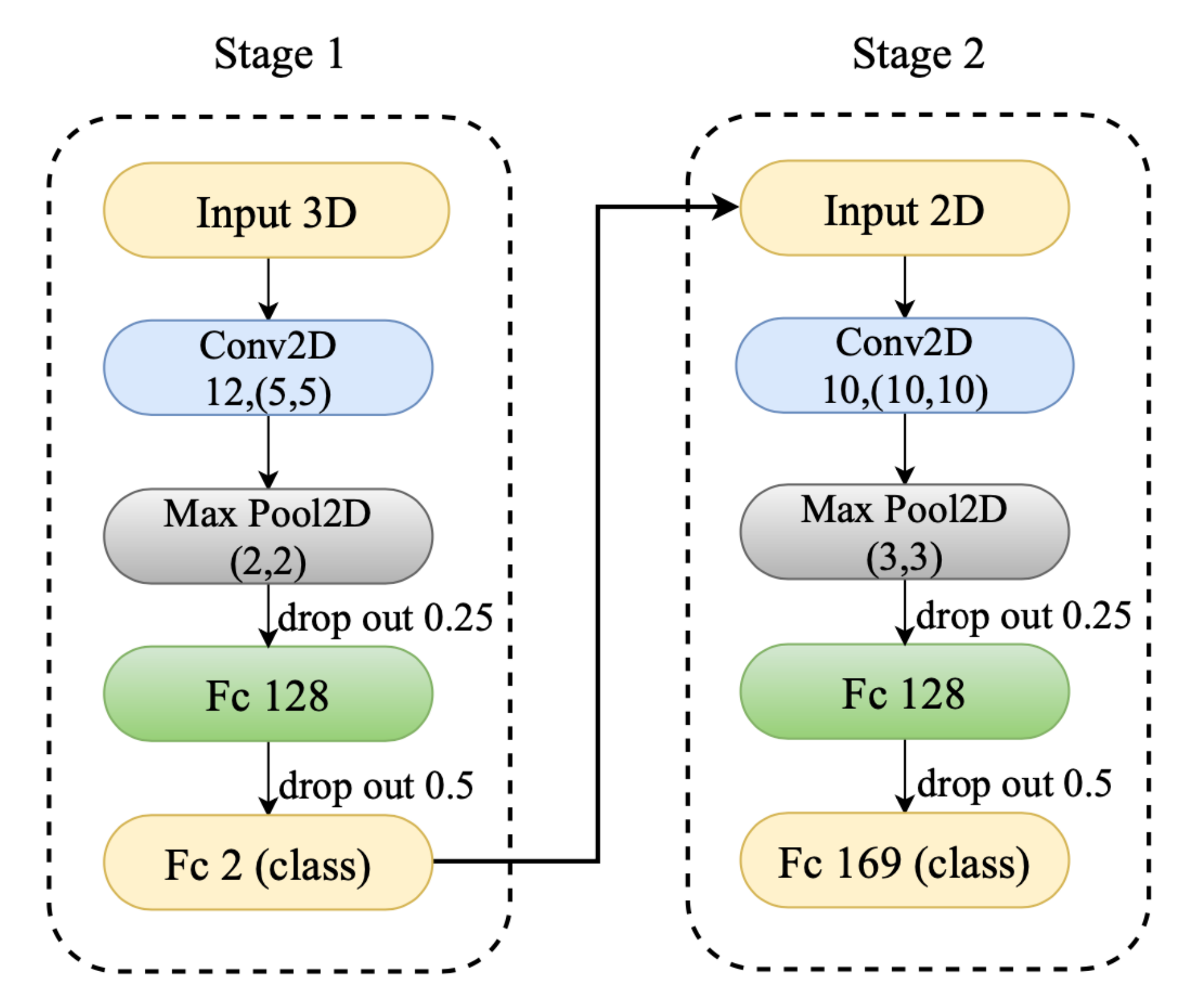}
    \caption{Model architecture for Stage 1 (detection) and Stage 2 (best beam pair prediction).}
    \label{fig:models}
\end{figure}

The network architecture after hyper-parameter tuning is shown in Fig.~\ref{fig:models}. The crops, which are RGB images, are passed to a two-dimensional convolutional layer with 12 filters of kernel size (5,5). The next layer is a max-pooling layer with the pool size of (2,2). After being flattened, the output is fed to a dense layer with 128 neurons. Finally, the output layer with two outputs is passed through the \textit{softmax} activation function for classification purposes. In order to prevent overfitting, we added two dropout layers after convolutional and dense layers with the rate of 0.25 and 0.5, respectively. Furthermore, in order to minimize the inference time, we intentionally searched for the simplest model embodiment that ensured the desired level of accuracy in our experiments.

Note that our designed binary classifier gets a $W\times W$ image as input and predicts the corresponding label, i.e. \textit{Background} or \textit{Antenna array}. Given an input images to our pipeline, first, the input image is cropped with the window size of $W$ and stride size of $S$, as described previously. Second, Each crop is fed to the trained binary classifier to decide if the input crop is background or not. If the predicted label of the window is \textit{Background} the entire window is mapped to 0; however, the \textit{Antenna array} window is mapped to 1. Finally, the decisions are put together, in the same order as crop generation, to create a \textit{bit map}. The resulting bit map will have the height $\lfloor\frac{H-W}{S}+1\rfloor$ and width $\lfloor\frac{L-W}{S}+1\rfloor$, according to \eqref{crops} and represents the location of transmitter and receiver in the image. We evaluate the performance of this stage in the Sec. \ref{Result:stage1}.
\subsection{Stage 2: Predicting Best Beam Pairs}
\label{second stage}
Using the binary classifier derived from the Stage 1, the bit map of the input image is generated and used as input to the second CNN to predict the labels, best beam configuration as described in table \ref{Tab:dataset}. Note that, while in the first stage the input is an RGB image with three channels, in the second stage each bit map has only one channel. We preserve the model structure from Stage 1 and adjust the hyperparameters as shown in Fig.~\ref{fig:models}. We increase the number of neurons in the classifier layer to 169, which is the number of possible beam combinations. It should be noted that while collecting data for wood and card box as the obstacle, only 18 and 20 out of 169 classes are emerged in the dataset collected in our experiments. We shuffle our expanded dataset on various light conditions and split it as $(75\%,15\%,15\%)$ to generate the train, validation, and test sets, respectively. Finally, we train the model to predict the best beam pair configuration at the transmitter and receiver side. The performance of this stage is assessed in section \ref{Result:stage2}. 

\subsection{Handling Camera Field of View Obstructions}
\label{Mutli-camera}
Since we use visual information for inferring the best beam pair, our prediction accuracy depends on how visible the transmitter and receiver devices are in the input image. In the case of obstructed view, multiple cameras can be deployed to reduce blind spots. Our algorithm can be trivially extended to collectively extract relevant features from multiple view angles and reinforce the performance.

In order to incorporate the information from different angles, first we use our proposed method in Stage 1 to infer the location of transmitter and receiver in the images taken from different angles, obtaining multiple bit maps. After generating the bit maps, we stack them in different channels and pass it to Stage 2 for inferring best beam pair. To adopt our proposed method for multiple camera case, we only need to change the input shape to the second stage and increase the number of channels to total number of cameras. We use our dataset collected with card box as obstacle to evaluate the performance of multiple camera deployment in section \ref{Results:blocked view}.

\subsection{Enhancements to CNN Architecture}
\label{transform to FCC}
In our proposed method, we create small crops from the input image and feed it to our classifier. As a result, in Stage 1, the model needs to predict the label for total number of \eqref{crops} crops that might not be time efficient. We employ two different approaches to decrease the inference time. First, we compress our model as much as possible to reduce the number of operations, as shown in Fig. \ref{fig:models}. Second, we convert our model to a fully convolutional network (FCN) by taking steps presented in Algorithm \ref{cnn-to-fcn}. This conversion allows us to slide the original model very efficiently across all possible spatial positions on the entire image, in a single forward pass. 
\begin{algorithm}[t]
\SetAlgoLined
\KwIn{Convolutional neural network} 
\KwOut{Equivalent fully convolutional network} 
\For{Dense layers in CNN}{
 \begin{algorithmic}
    \STATE $d_w^{in}$ $\leftarrow$ \text{Get size of \textit{Dense} input;}\\
    \STATE $d_w^{out}$ $\leftarrow$ \text{Get size of \textit{Dense} output;}\\
    \STATE $W_d$ $\leftarrow$ \text{Get weights of \textit{Dense} layer;}\\
  \IF{First \textit{Dense} layer after \textit{Flatten} layer}
    \STATE $(f_w^{in},f_h^{in},f_d^{in})$ $\leftarrow$ \text{Get shape of \textit{Flatten} input;}\\
    \STATE $W_{d}^{\prime}$ $\leftarrow$ \text{Reshape $W_d$ from $(d_w^{in},d_w^{out})$ }\\
    \qquad\text{to  $(f_w^{in},f_h^{in},f_d^{in},d_w^{out})$;}\\
    \STATE Remove \textit{Flatten} layer;
    \STATE Convert \textit{Dense} to \textit{Conv2D} layer 
    with $f_d^{in}$ filters \\ 
    \qquad of size $(f_w^{in},f_h^{in})$ and load new weights $W_{d}^{\prime}$; 
  \ELSIF{Other \textit{Dense} layer}
    \STATE $W_{d}^{\prime}$ $\leftarrow$ Reshape $W_d$ from $(d_w^{in},d_w^{out})$ \\
    \qquad to  $(1,1,d_w^{in},d_w^{out})$;\\
    \STATE \text{Convert \textit{Dense} to \textit{Conv2D} layer with $d_w^{out}$ filters} \\
    \STATE \qquad \text{of size $(1,1)$ and load new weights $W_{d}^{\prime}$;} \\
  \ENDIF
\end{algorithmic} 
 }
\caption{Convert CNN to FCN}
\label{cnn-to-fcn}
\end{algorithm}
Although this transformation does not eliminate the need for training on crops, it can speed up the prediction speed while testing. We evaluate the performance of both original and fully convolutional architectures in section \ref{Result:prediction time}.
\section{Performance Evaluation}
\begin{figure*}[t!!]
    \centering
    \begin{subfigure}{0.45\textwidth}
    \centering
    \includegraphics[scale=0.22]{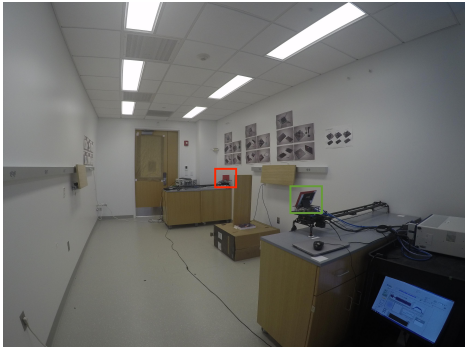}
    \vspace{0.22cm}
    \caption{}
    \label{fig8_original}
    \end{subfigure}
    \begin{subfigure}{0.45\textwidth}
    \centering
    \includegraphics[scale=0.26]{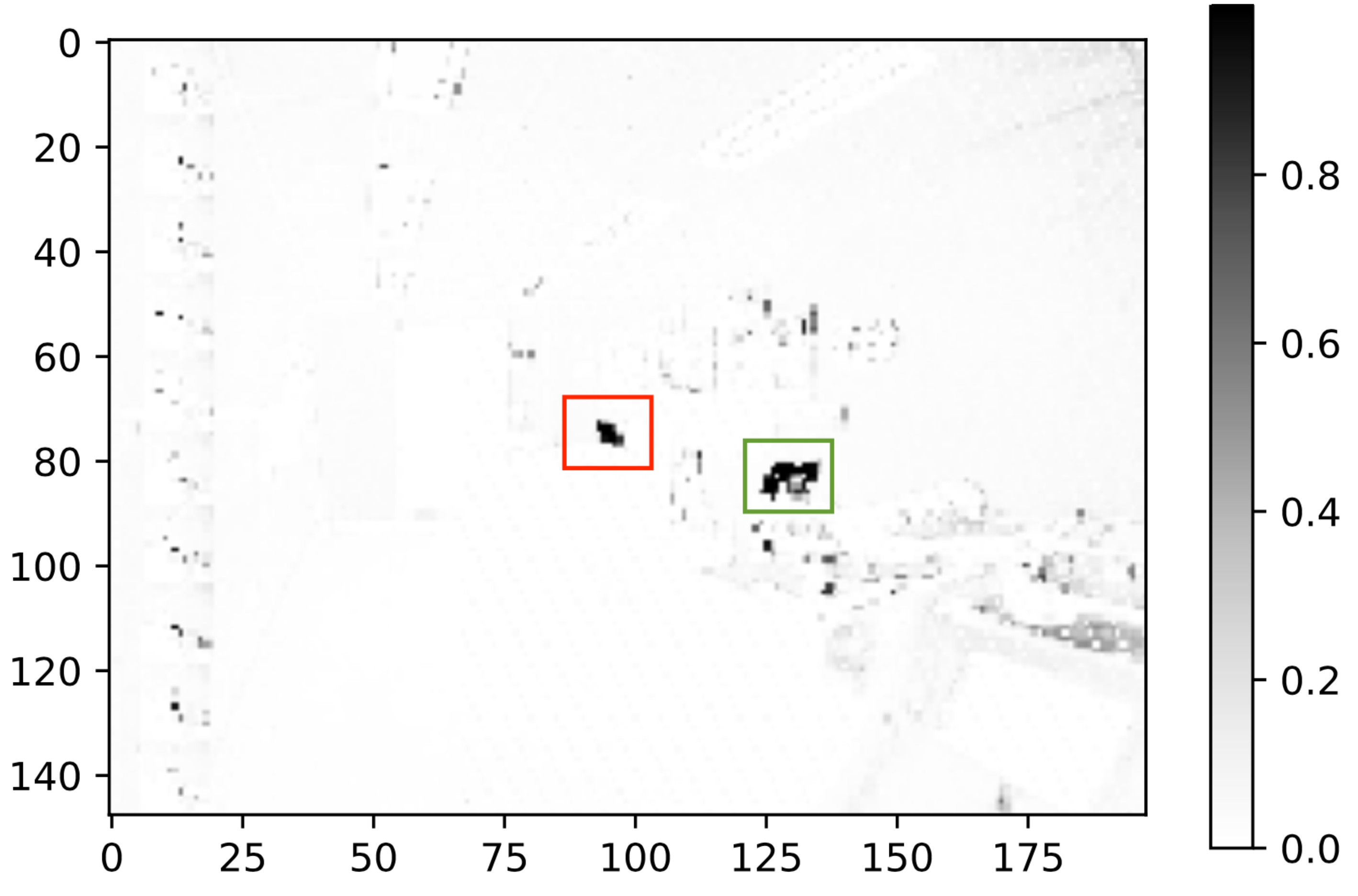}
    \caption{}
    \label{fig8_probabilty}
    \end{subfigure}
    \caption{The input image (a) of case (1,5) passes through Stage 1 in our design. The heat map (b) is generated using our binary classifier and shows the location of transmitter and receiver. }
    \label{bitmap and original}
\end{figure*}

\label{results}
In this section, we will provide the results of our proposed method on the dataset described in section \ref{sec:expSetup}. We used Keras 2.1.6 on top of Tensorflow backend (version 1.9.0) to implement and train the classifiers.
\subsection{Stage 1 (Detection)}
\subsubsection{Binary Classifier Accuracy}
\label{Result:stage1}
We resize the original RGB images from the camera with shape $(3000,4000,3)$ to $(750,1000,3)$ and use it as input of our pipeline. The first step is to train our binary classifier on $W\times W$ samples of \textit{Antenna array} and \textit{Background}. The window size needs to be large enough to extract useful information from the crops, and small enough at the same time to differentiate two adjacent cases. We empirically determine the window size of 12 and stride size of 5 for our experiment. The dataset for Stage 1 includes 732603 and 733903 cropped samples of \textit{Background} and \textit{Antenna array}, respectively, and the binary classifier achieves the accuracy of $99\%$ on the test set, demonstrating effective separation of the antenna arrays from the background in the input image.

Consider case (1,5) as an example, (see Fig.~\ref{fig8_original}). The output of the Stage 1 binary classifier is a prediction matrix with the shape of (29304,2), i.e. number of crops dervied from \eqref{crops} and number of classes. Each row represents the probability of belonging to \textit{Background} or \textit{Antenna array} class for the corresponding crop. Fig.~\ref{fig8_probabilty} shows the heat map of prediction probability for \textit{Antenna array} class. In this figure, the brightness of each pixel decreases as the crop has a higher prediction probability for our class of interest, i.e. \textit{Antenna array}. We separate the top 60 candidates for the \textit{Antenna array} class and arrange them in the same order we cropped the image and create a 2D bit map with the shape (148,198) as described in \eqref{crops}. Although the output feature maps present some mis-classifications, we note that it does not have a major impact on the system performance.  

\subsubsection{Intersection Over Union}
Each bit map can be interpreted as a set of points with two major clusters, representing the location of transmitter and receiver. We use Intersection over Union (IoU) metric to assess the performance of Stage 1, i.e. \textit{Detection}. While detecting an object in an image, the ground truth area is referred to a rectangle around the object of the interest which contains the entire object, denoted as $B_{gt}$. A predictor, a CNN for instance, is then used to estimate the location of the object in the image. A rectangle around predictor estimated pixels denotes the detector prediction for the object location, $B_{p}$. The IoU evaluates the object detection accuracy and defined as:
\begin{equation}
    IoU = \frac{Area\{B_{p} \cap B_{gt}\}}{Area\{B_{p} \cup B_{gt}\}}.
\end{equation}

\begin{figure}[t]
    \centering
    \includegraphics[scale=0.4]{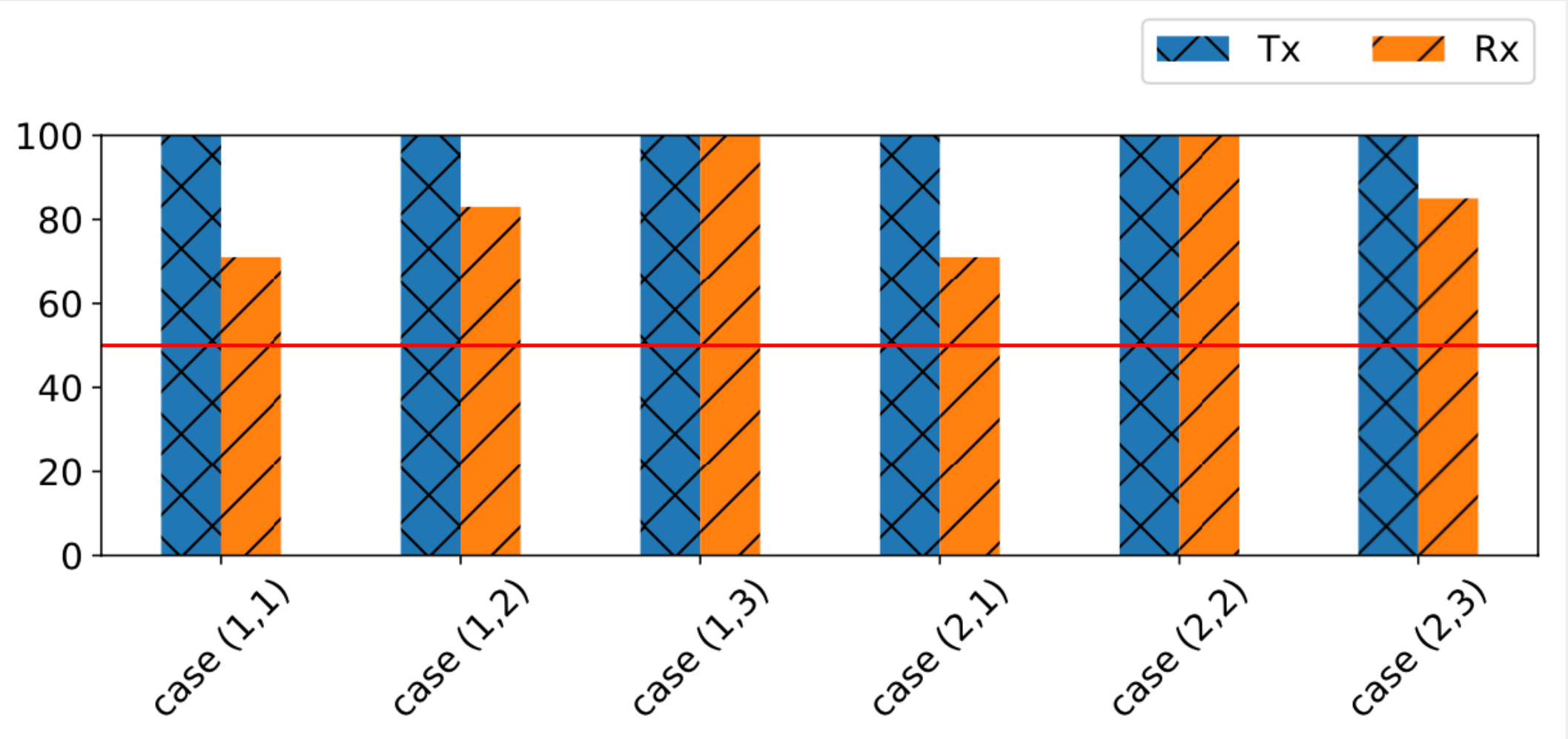}
    \caption{Intersection over Union metric for six cases from dataset.}
    \label{barplot}
\end{figure}

In order to measure the detection area for each bitmap, we extract the index of non-zeros elements and find the centroid of the transmitter and receiver clusters. We draw a rectangle around the centroids and increase its dimensions by one pixel in each iteration. We stop when there is no point to be added to the rectangle. Fig.~\ref{barplot} shows the IoU metric for the transmitter and receiver localization over 6 different scenarios. When IoU exceeds 0.5, the object detection is accomplished, also known as true positive. On the other hand, detection fails with a false positive outcome when IoU is below 0.5. From Fig.~\ref{barplot}, we see that the IoU is higher than the 0.5 threshold, for all cases. Thus, we conclude that our proposed algorithm successfully tracks the relative location of the transmitter and receiver. 

\subsection{Stage 2 (Beam Classifier Accuracy)}
\label{Result:stage2}
The structure of the dataset for Stage 2 contains the bitmaps, for all 1250 cases in our expanded dataset, and the associated best beam pair as presented in table \ref{Tab:dataset}. In our setting, the labels are tuples depicting the best beam pair at the transmitter and receiver. In order to adapt them for training, we map each pair to a unique number and then apply one-hot encoding on new labels. By following the instruction provided in section \ref{second stage}, we divide the dataset into $(75\%,15\%,15\%)$ and train our model, shown in Fig.~\ref{fig:models}, for 10 epochs. Our designed classifier achieves $99\%$ accuracy while predicting the best beam pairs on the test set. For both stages, we use batch size of 256 and Adam optimizer with a learning rate of 0.001.
\subsection{Handling Transceiver View Obstruction}
\label{Results:blocked view}
\begin{figure}[ht]
    \centering
    \begin{subfigure}{0.23\textwidth}
    \centering
    \includegraphics[scale=0.155]{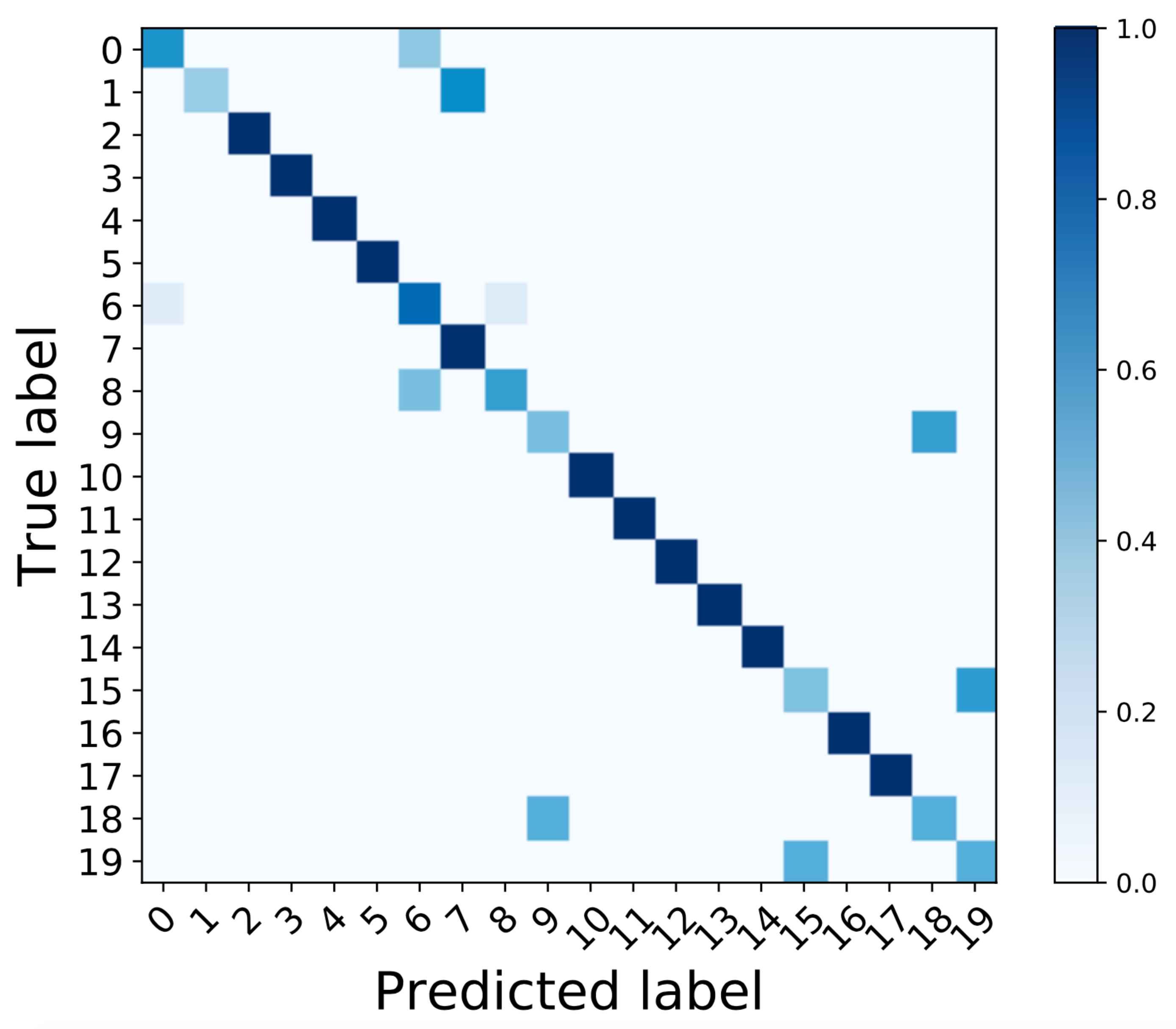}
    \caption{}
    \label{blocked}
    \end{subfigure}
    \begin{subfigure}{0.23\textwidth}
    \centering
    \includegraphics[scale=0.155]{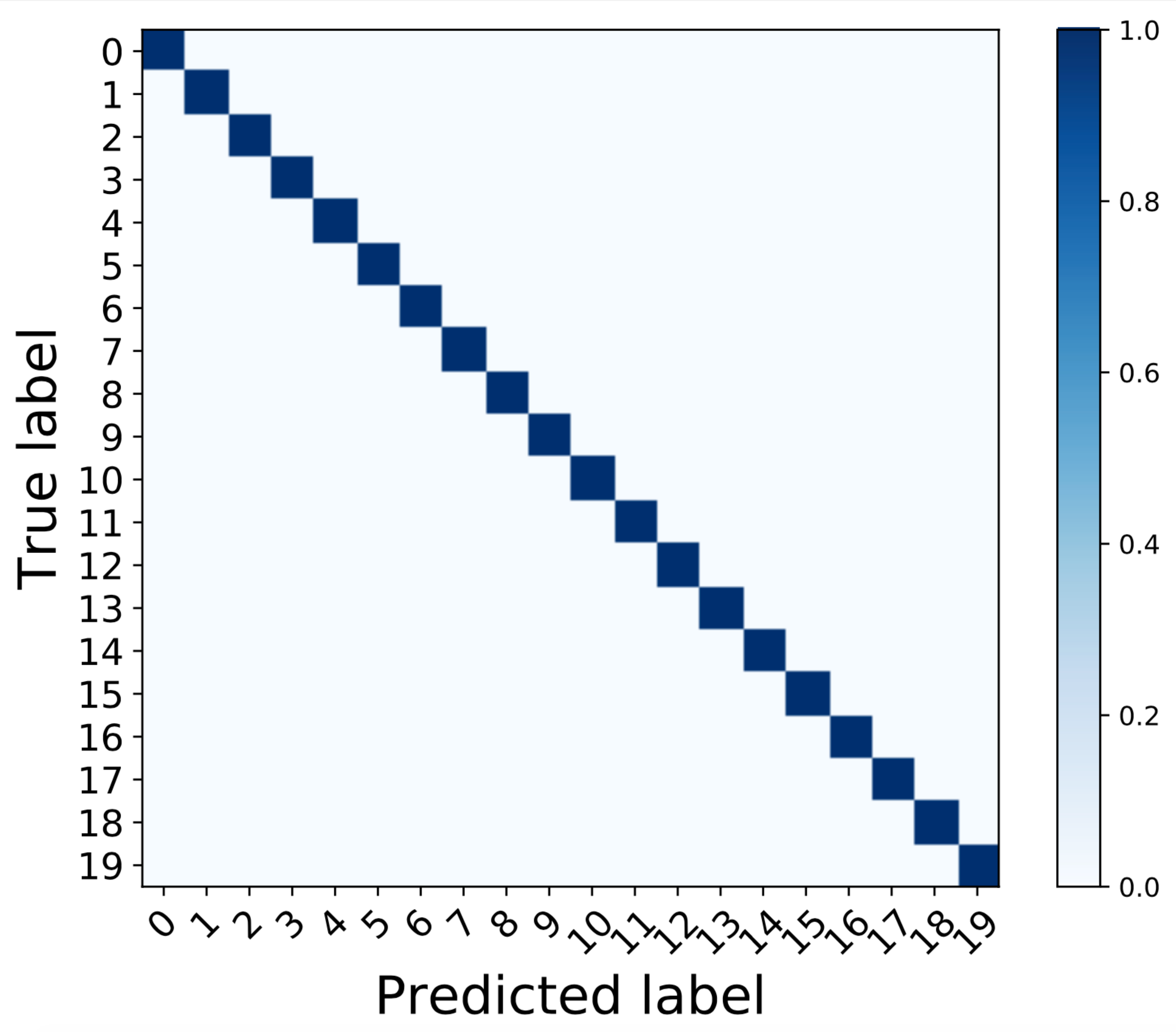}
    \caption{}
    \label{stacked}
    \end{subfigure}
    \caption{Confusion matrix for using (a) single bit map from blocked angle; (b) stacked bit maps from both angles.}
    \label{confusion matrix}
\end{figure}

In our testbed, the first camera is positioned such that it has a clear 
view of the transmitter and receiver while the second camera has a blocked view of the transmitter for 250 cases out of 1250 cases included in the dataset. We observed a drop from $99\%$ to $80\%$ accuracy while switching from the first to second angle. 
Fig.~\ref{confusion matrix} shows the confusion matrix on best beam pair estimation for the blocked angle and the improvement achieved by using multiple cameras, as proposed in section \ref{Mutli-camera}. Our experiment shows that the accuracy reaches back to $99\%$ by stacking the bitmap of different angles. Thus, we can use multiple cameras to compensate for the blocked angles. 

\subsection{Prediction Time}
\label{Result:prediction time}
\begin{figure}[t]
    \centering
    \includegraphics[width=0.45\textwidth]{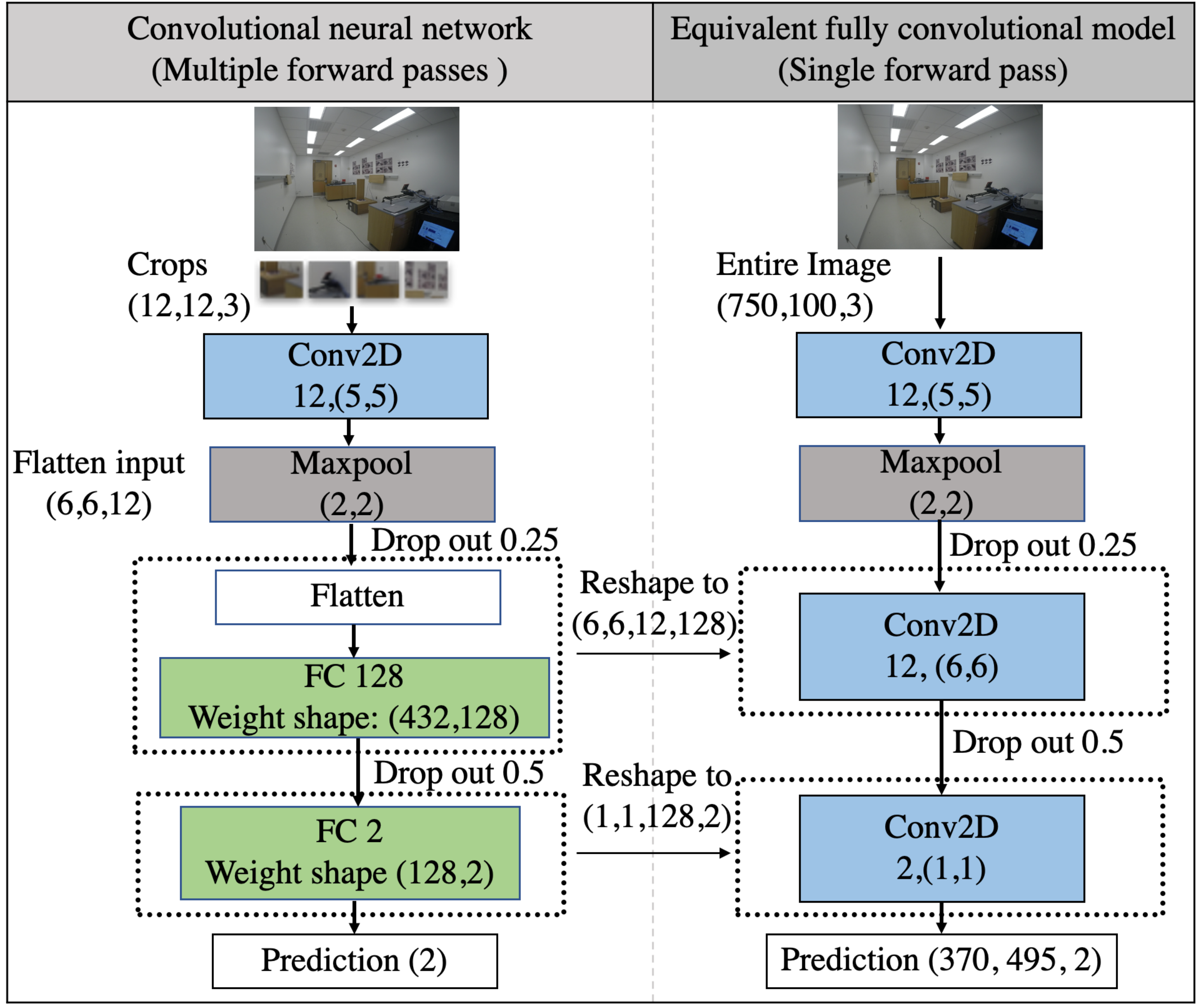}
    \caption{Transferring trained model in Stage 1 to a fully convolutional architecture to speed up inference using algorithm \ref{cnn-to-fcn}. The equivalent fully convolutional network generates the bit map in a single forward pass, while the original model needs to predict the label \eqref{crops} times.}
    \label{fig:fcn}
\end{figure}

Fig.~\ref{fig:fcn} denotes the original and transformed model, derived from Algorithm \ref{cnn-to-fcn}, for Stage 1. We observe that the equivalent FCN passes the entire image in a single forward path and generates a single (370,495,2) prediction matrix.
In order to evaluate the inference speed, we pass a single image hundred times through our pipeline and measure the prediction time by setting a timer and subtracting the time stamp before and after prediction. We report the average prediction time taken over all samples as the required time for prediction. The NVIDIA V100 GPU with 32GB memory is used to run the experiments. 

We reduce the prediction time from 4.47s to 2.0544ms in Stage 1 by converting our model to a fully convolutional one, explained in section \ref{transform to FCC}. For Stage 2, the conversion does not bring any benefit in terms of computing time as we evaluate a single input and produce a single output. So, we keep the initial structure with 1.05ms prediction time. Accordingly, our proposed method predicts the best beam pair in 3.104ms, approximately. This outperforms other approaches in Table \ref{compare different devices} by 93\% reduction in time taken for beam alignment, considering the same number of possible codebook configurations.

\section{Conclusion}
In this paper, we introduced the concept of using visual information as an alternative for exhaustive beam sweeping algorithm proposed by 802.11ad standard. We proposed a two-stage approach to extract the location of transmitter and receiver from the images and map them to the best beam pairs. We validated our method on a real-world dataset collected using National Instruments mmWave transceiver. Our method can predict the best beam pair with $99\%$ accuracy in 3.104ms for the hardware used in the testbed.
\section{Acknowledgements}
The authors gratefully acknowledge support by the National Science Foundation (grant CCF-1937500).
\bibliographystyle{IEEEtran}
\bibliography{reference}

\begin{thebibliography}{10}
\providecommand{\url}[1]{#1}
\csname url@samestyle\endcsname
\providecommand{\newblock}{\relax}
\providecommand{\bibinfo}[2]{#2}
\providecommand{\BIBentrySTDinterwordspacing}{\spaceskip=0pt\relax}
\providecommand{\BIBentryALTinterwordstretchfactor}{4}
\providecommand{\BIBentryALTinterwordspacing}{\spaceskip=\fontdimen2\font plus
\BIBentryALTinterwordstretchfactor\fontdimen3\font minus
  \fontdimen4\font\relax}
\providecommand{\BIBforeignlanguage}[2]{{%
\expandafter\ifx\csname l@#1\endcsname\relax
\typeout{** WARNING: IEEEtran.bst: No hyphenation pattern has been}%
\typeout{** loaded for the language `#1'. Using the pattern for}%
\typeout{** the default language instead.}%
\else
\language=\csname l@#1\endcsname
\fi
#2}}
\providecommand{\BIBdecl}{\relax}
\BIBdecl

\bibitem{6732923}
S.~{Rangan}, T.~S. {Rappaport}, and E.~{Erkip}, ``Millimeter-wave cellular
  wireless networks: Potentials and challenges,'' \emph{Proceedings of the
  IEEE}, vol. 102, no.~3, pp. 366--385, 2014.

\bibitem{roh2014millimeter}
W.~Roh, J.-Y. Seol, J.~Park, B.~Lee, J.~Lee, Y.~Kim, J.~Cho, K.~Cheun, and
  F.~Aryanfar, ``Millimeter-wave beamforming as an enabling technology for 5g
  cellular communications: Theoretical feasibility and prototype results,''
  \emph{IEEE communications magazine}, vol.~52, no.~2, pp. 106--113, 2014.

\bibitem{assasa2019enhancing}
H.~Assasa, J.~Widmer, T.~Ropitault, and N.~Golmie, ``Enhancing the ns-3 ieee
  802.11 ad model fidelity: Beam codebooks, multi-antenna beamforming training,
  and quasi-deterministic mmwave channel,'' in \emph{Proceedings of the 2019
  Workshop on ns-3}, 2019, pp. 33--40.

\bibitem{yaman2016reducing}
Y.~Yaman and P.~Spasojevic, ``Reducing the los ray beamforming setup time for
  ieee 802.11 ad and ieee 802.15. 3c,'' in \emph{MILCOM 2016-2016 IEEE Military
  Communications Conference}.\hskip 1em plus 0.5em minus 0.4em\relax IEEE,
  2016, pp. 448--453.

\bibitem{hou2019prediction}
Z.~Hou, C.~She, Y.~Li, L.~Zhuo, and B.~Vucetic, ``Prediction and communication
  co-design for ultra-reliable and low-latency communications,'' \emph{IEEE
  Transactions on Wireless Communications}, vol.~19, no.~2, pp. 1196--1209,
  2019.

\bibitem{niwebsite}
\BIBentryALTinterwordspacing
5g nr for wireless communications. [Online]. Available: \url{http://ni.com/5g}
\BIBentrySTDinterwordspacing

\bibitem{8101513}
M.~{Hashemi}, C.~E. {Koksal}, and N.~B. {Shroff}, ``Out-of-band millimeter wave
  beamforming and communications to achieve low latency and high energy
  efficiency in 5g systems,'' \emph{IEEE Transactions on Communications},
  vol.~66, no.~2, pp. 875--888, 2018.

\bibitem{nitsche2015steering}
T.~Nitsche, A.~B. Flores, E.~W. Knightly, and J.~Widmer, ``{Steering with eyes
  closed: mm-wave beam steering without in-band measurement},'' in \emph{2015
  IEEE Conference on Computer Communications (INFOCOM)}.\hskip 1em plus 0.5em
  minus 0.4em\relax IEEE, 2015, pp. 2416--2424.

\bibitem{ali2017millimeter}
A.~Ali, N.~Gonz{\'a}lez-Prelcic, and R.~W. Heath, ``Millimeter wave
  beam-selection using out-of-band spatial information,'' \emph{IEEE
  Transactions on Wireless Communications}, vol.~17, no.~2, pp. 1038--1052,
  2017.

\bibitem{gonzalez2016radar}
N.~Gonz{\'a}lez-Prelcic, R.~M{\'e}ndez-Rial, and R.~W. Heath, ``Radar aided
  beam alignment in mmwave v2i communications supporting antenna diversity,''
  in \emph{2016 Information Theory and Applications Workshop (ITA)}.\hskip 1em
  plus 0.5em minus 0.4em\relax IEEE, 2016, pp. 1--7.

\bibitem{va2017inverse}
V.~Va, J.~Choi, T.~Shimizu, G.~Bansal, and R.~W. Heath, ``Inverse multipath
  fingerprinting for millimeter wave v2i beam alignment,'' \emph{IEEE
  Transactions on Vehicular Technology}, vol.~67, no.~5, pp. 4042--4058, 2017.

\bibitem{wang2018mmwave}
Y.~Wang, M.~Narasimha, and R.~W. Heath, ``Mmwave beam prediction with
  situational awareness: A machine learning approach,'' in \emph{2018 IEEE 19th
  International Workshop on Signal Processing Advances in Wireless
  Communications (SPAWC)}.\hskip 1em plus 0.5em minus 0.4em\relax IEEE, 2018,
  pp. 1--5.

\bibitem{alkhateeb2018machine}
A.~Alkhateeb, I.~Beltagy, and S.~Alex, ``Machine learning for reliable mmwave
  systems: Blockage prediction and proactive handoff,'' in \emph{2018 IEEE
  Global Conference on Signal and Information Processing (GlobalSIP)}.\hskip
  1em plus 0.5em minus 0.4em\relax IEEE, 2018, pp. 1055--1059.

\bibitem{8792137}
T.~{Nishio}, H.~{Okamoto}, K.~{Nakashima}, Y.~{Koda}, K.~{Yamamoto},
  M.~{Morikura}, Y.~{Asai}, and R.~{Miyatake}, ``Proactive received power
  prediction using machine learning and depth images for mmwave networks,''
  \emph{IEEE Journal on Selected Areas in Communications}, vol.~37, no.~11, pp.
  2413--2427, 2019.

\bibitem{8395149}
A.~{Alkhateeb}, S.~{Alex}, P.~{Varkey}, Y.~{Li}, Q.~{Qu}, and D.~{Tujkovic},
  ``Deep learning coordinated beamforming for highly-mobile millimeter wave
  systems,'' \emph{IEEE Access}, vol.~6, pp. 37\,328--37\,348, 2018.

\bibitem{Terragraph}
``{Terragraph by Facebook. Solving the Urban Bandwidth Challenge},''
  \url{https://terragraph.com}, accessed: 2020-05-31.

\end{thebibliography}
\end{document}